\documentclass[aps,prb,10pt,showpacs,showkeys,preprintnumbers,superscriptaddress,a4paper,twocolumn]{revtex4-1}
\usepackage{amsmath,amsthm,amssymb}
\usepackage[utf8]{inputenc}
\usepackage[T1]{fontenc}
\usepackage{textcomp}
\usepackage{graphicx}
\usepackage[export]{adjustbox}
\usepackage[caption=false]{subfig}
\usepackage{lmodern}
\DeclareUnicodeCharacter{B0}{\textdegree}
\usepackage{placeins}

\usepackage{amstext}
\usepackage{lipsum}
\usepackage{enumitem}
\usepackage{mwe}
\usepackage{mathtools}
\usepackage{afterpage}
\usepackage{epsfig}
\usepackage{epstopdf} 
\usepackage{filecontents}
\usepackage[hyperindex,pdftex,colorlinks,citecolor=blue,filecolor=blue,linkcolor=blue,urlcolor=blue]{hyperref}

\numberwithin{equation}{figure}

\def\beq{\begin{equation}}
\def\eeq{\end{equation}}
\def\bea{\begin{eqnarray}}
\def\eea{\end{eqnarray}}
\def\ba{\begin{array}}
\def\ea{\end{array}}

\begin{document}
\title{Dipole-dipole interaction induced phases in Hydrogen-bonded squaric acid crystal}
\author{Vikas Vijigiri}\email{vikasvikki@iopb.res.in}\affiliation{Institute of Physics, Bhubaneswar-751005, Orissa, India}
\affiliation{Homi Bhabha National Institute, Mumbai - 400 094, Maharashtra, India}
\author{Saptarshi Mandal }\email{saptarshi@iopb.res.in}\affiliation{Institute of Physics, Bhubaneswar-751005, Orissa, India}
\affiliation{Homi Bhabha National Institute, Mumbai - 400 094, Maharashtra, India}

\begin{abstract}
We study analytically the finite-temperature phase diagram of proton ordering of a quasi-two dimensional hydrogen-bonded system, namely the squaric acid crystal($\text{H}_2\text{C}_4\text{O}_4$). We take into account the four-spin interaction model at the zeroth order. Using an improvised loop algorithm within the Stochastic series expansion quantum monte carlo method, we find two distinct phases as we increase the temperature and magnetic-field.  One of the phase is the $\Pi_f$, the phase with long range ferroelectric order and the other being an intermediate state with strong local correlations, i.e, a quantum liquid-like state $\Pi_{ql}$. The transition to $\Pi_{f}$ shows a very small anomalous peak in the specific heat with strong dependence of critical temperature on the strength of dipole-dipole interaction. The presence of the small peak is attributed to the absence of macroscopic degeneracy in the presence of dipole-dipole interaction and re-entrance of such degeneracy to some extent at small temperature.  Though the  degenerate ground state manifold is identical for a four-spin interaction or appropriate two-spin interaction model at zeroth order, we find that for the former case, the required strength of dipole-dipole interaction is quite larger to induce a ferroelectric phase.
The work also presents an intricate connection of quantum fluctuation and thermal fluctuation in the presence of competing interaction with entropic effects.

\end{abstract}

\date{\today}

\maketitle
\section{Introduction}
\indent 

The observation of finite temperature phase transition in condensed matter system is a common phenomena since unknown past.
We know how various types of phase transition phenomenon, few like the metallic to superconducting phase transition or solid to liquid to vapor
transition or the structural or magnetic transition is brought in by the changes of temperature. As the temperature changes, the
interactions between constituent particles are renormalized or tuned so that new phases of matters appear. The interplay of  
fundamental interactions, quantum mechanics as well as the thermal fluctuation all play equally vital role in such event. In this 
article we will be discussing the composite effect of finite temperature along with the magnetic field phase diagram of a well known material known as squaric acid, commonly represented
as $\text{H}_2 (\text{SQ})_4$ where $\text{SQ}$ generally represent a structural unit(e.g: $\text{C}_4\text{O}_4$).  This particular material had drawn the attention of condensed
matter physicists (theory and experiment alike) over more than half a century ~\cite{R.Blinc,E.Matsushita,J.C.Slater,G.A.Samar,
P.S.Peerc,P.G.deGennes,J.D.Bernal}. The material is very intriguing, as the primary interaction is governed 
by the hydrogen bonds and the associated proton dynamics. The material is a layered three dimensional one where each 2D layer is made of square lattice 
like structure\cite{E.J.Samuelsen} with each square containing a unit of $\text{SQ}$. Each of them are held together by Hydrogen bonds which is 
mediated by protons. An individual proton undergoes quantum motions between double well potential\cite{R.Blinc,E.Matsushita} as found
from the isotopic study of the material. The proton motion is conveniently modelled as  quantum spins(spin 1/2). \\
\indent $\text{H}_2\text{SQ}$ is very similar to water ice system, where  the local constraint in the form of "ice-rules" dictating the proton ordering 
restricts the ground state to have configurations where in each molecule having exactly four hydrogen-bonds with two protons being near 
to $\text{C}_4\text{O}_4$ unit and the other two being at far\cite{L.Pauling,R.Savit,HD.Maier,G.F.Reiter}. Apart from the proton 
dynamics and the interactions between them, the material posses dipole-dipole interactions which is evident from the fact that the material 
undergoes a paraelectric to ferroelectric phase transition\cite{E.J.Samuelsen}. These systems are of particular interest since the last decade 
because not only that these systems exhibit some of the peculiar properties: showing no signs of phase transition, i.e, they tend to be 
disordered down to few milli Kelvins, there is frustration or competition leading to "residual" or "zero-point entropy"\citep{L.Pauling}. 
In some cases, strong fluctuations with high degree of entanglement between the ice-rule configurations may lead to exotic phases 
like quantum-spin liquid~\citep{N.Shannon1,N.Shannon,K.A.Ross,O.F.Syljuasen,E.Ardonne}.  Long back, this "ice-rule" constraint and
the movement of protons on  $\text{H}_2\text{SQ}$  were mapped  into suitable vertex model or quantum pseudo spin model \citep{U.Deininghaus,J.F.Stilck,V.I.Zinenko} to understand the  antiferroelectric properties. Further, the thermal motion of a $\text{SQ}$-unit in a squaric acid system is strongly coupled
with the protons movements. This motivated the study of interaction of phonons with pseudo-spins~\citep{E.Matsushita1,D.Semmingsen,B.K.Chaudhuri}.
 Though these pioneering works had successfully examined some key features of the system, many of the theoretical methods were mostly limited to the 
mean-field level without totally considering the effect of geometrical frustration more rigorously. For example, quantum fluctuation along with the entropic effects, which is presumably important for understanding the quantum  para-electricity under pressure has not been seriously considered so far.\\
\begin{figure}[htp]
\includegraphics[width=8.6cm,height=5.0cm]{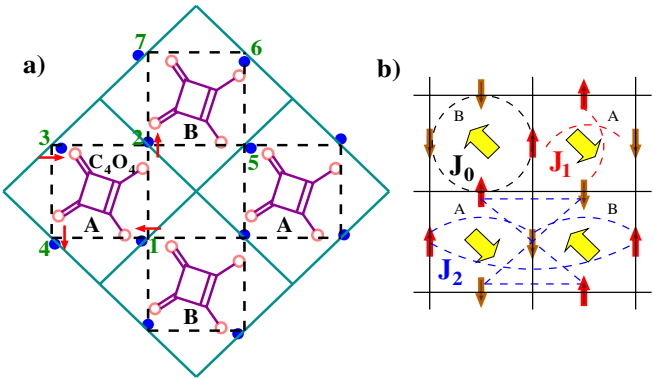}
\caption{\label{square}a) Squaric Acid network system, the blue dots represent the hydrogen atoms and the small pink circles represent the oxygen 
atoms of the molecule $\text{H}_2\text{SQ}$ forming a quasi-2D configuration of hydrogen-bonded network, here $\text{SQ}$ being the  $\text{C}_4\text{O}_4$ 
unit is shown. The dashed brown lines show the physical lattice and the solid cyan lines form the lattice on which the model is built on. The indices 
shown represent the corresponding spins as shown in in Eq.(\ref{Hamiltonian}). (b) We present one of the configuration of the ferroelectric ordering, 
the yellow arrow indicate the polarization axis. The dotted color lines are drawn to indicate bonds/plaquette for each interaction of the Hamiltonian, $
J_0$(black), $J_0$(red), $J_2$(blue). }       
\end{figure} 
\indent As already mentioned, hydrogen-bonded(HB) crystals are also known for their ferroelectric polarizability due to certain proton 
ordering(e.g: $\text{KH}_2\text{PO}_4$, $\text{H}_2\text{SQ}$). There has been ample studies regarding the electronic polarization 
emerging from certain ordering of proton configurations in the system with emphasis on both the fundamental physics and the application 
to electronic devices\cite{M.E.Lines,S.Horiuchi}. Apart from these, One of the interesting feature of $\text{H}_2\text{SQ}$ is that, 
in general, the external pressure increases the tunneling rate of protons between the two potential minima, which reduces local polarizations, 
and consequently, suppresses the ferroelectricity. Indeed, in H2SQ, $\text{T}_\text{c}$ is suppressed with increasing pressure. 
A peculiar intermediate state, however, appears before the polarization is lost in each molecule: the macroscopic polarization vanishes 
while the polarization in each molecule is retained\citep{Y.Moritomo}. Consequently, a quantum paraelectric state is realized in the 
low-temperature limit. Thus the material presents an unique test bed where the effect of external pressure as well as temperature
could be investigated on the proton dynamics so that a number of interesting and competing aspect can be observed in a single study. 
Here we consider the model Hamiltonian\eqref{Hamiltonian} given below. There are three competing interactions at the effective order. The model Hamiltonian at zeroth order consists of a four-spin plaquette interaction with a strength $J_0$. 
When the Hamiltonian is studied in the presence of a magnetic field characterized by the parameter $K$, the system shows a confinement-deconfinement phase transition. The model also includes a next-nearest-neighbor Ising-like interaction with a strength $J_1$ and a dipole-dipole interaction $J_2$<$J_1$. Usually, the presence of $J_2$ causes the ferroelectricity of these materials. The model was shown to exhibit both confinement-deconfinement transition (CDT) and ferroelectric quantum phase transition (FT) for an appropriate set of parameters of the model Hamiltonian. However, in the present paper we study the finite temperature (incorporated with magnetic-field) properties of each of these phases. In the presence of intra-molecular coupling term $J_1$, we identify a smooth crossover to paraelectric phase from a globally disordered dipole phase. In the presence of $J_2$, the system gets ferroelectrically ordered at low temperatures and with the increase of field strength $K$, the ferroelectric ordered is being destroyed to a paraelectric phase which also undergoes a finite temperature second order phase transition to a conventional paraelectric phase. We then numerically chart out the phase diagram in the $T-K$ plane. In particular to this model Hamiltonian unlike the earlier studies,  which  were limited at the mean-field level, the effect of geometrical frustration has not been fully clarified yet. We use a unbiased stochastic series expansion(SSE) quantum monte carlo(QMC) technique to take into account the quantum fluctuation under the entropic effects(arising from the ice-rules), which is presumably important for understanding the quantum paraelectricity under pressure. Our study is also important in a way that for such an interacting four-spin Hamiltonian, we have been able to obtain the phase diagram successfully with appropriate quantum monte carlo scheme.\\ 
\indent The organization of this paper is as follows. In Sec.\ref{mod-meth}, we introduce the effective model Hamiltonian and methods. After introducing the pseudo-spin model in Sec.\ref{phys-quant}, we briefly discuss the  implementation scheme of SSE QMC method used for our Hamiltonian in the subsection of method part. We then explain the order parameter that has been calculated in our analysis. The results obtained from our SSE QMC simulation has been presented in details in Sec.\ref{results-qmc}. In Sec.\ref{disc}, we discuss our results in a broader perspective including a qualitative comparison with the relevant earlier experimental and theoretical study. We summarize our results in Sec.\ref{summ}.
\section{Model and method}
\label{mod-meth}
\subsection{Pseudo-spin Model}
In this section we introduce the full Hamiltonian that we consider for our study. The terms that we included has already been discussed before as well \cite{Chern,Vikas} and is given below
\begin{eqnarray}
\label{Hamiltonian}
H &=& -J_0\sum_{\langle ijkl \rangle} A_{\square} + J_1\sum_{\langle ij \rangle}B_{\square} \nonumber \\
&& - J_2\sum_{\langle AB \rangle}\vec{P}_{A}\cdot\vec{P}_{B} -K\sum_{i}\sigma_{i}^{x}
\end{eqnarray}
\begin{figure}[htp]
\hspace{3.0cm}
\centering
\label{cluster}
\includegraphics[width=8.5cm,height=4.5cm,left]{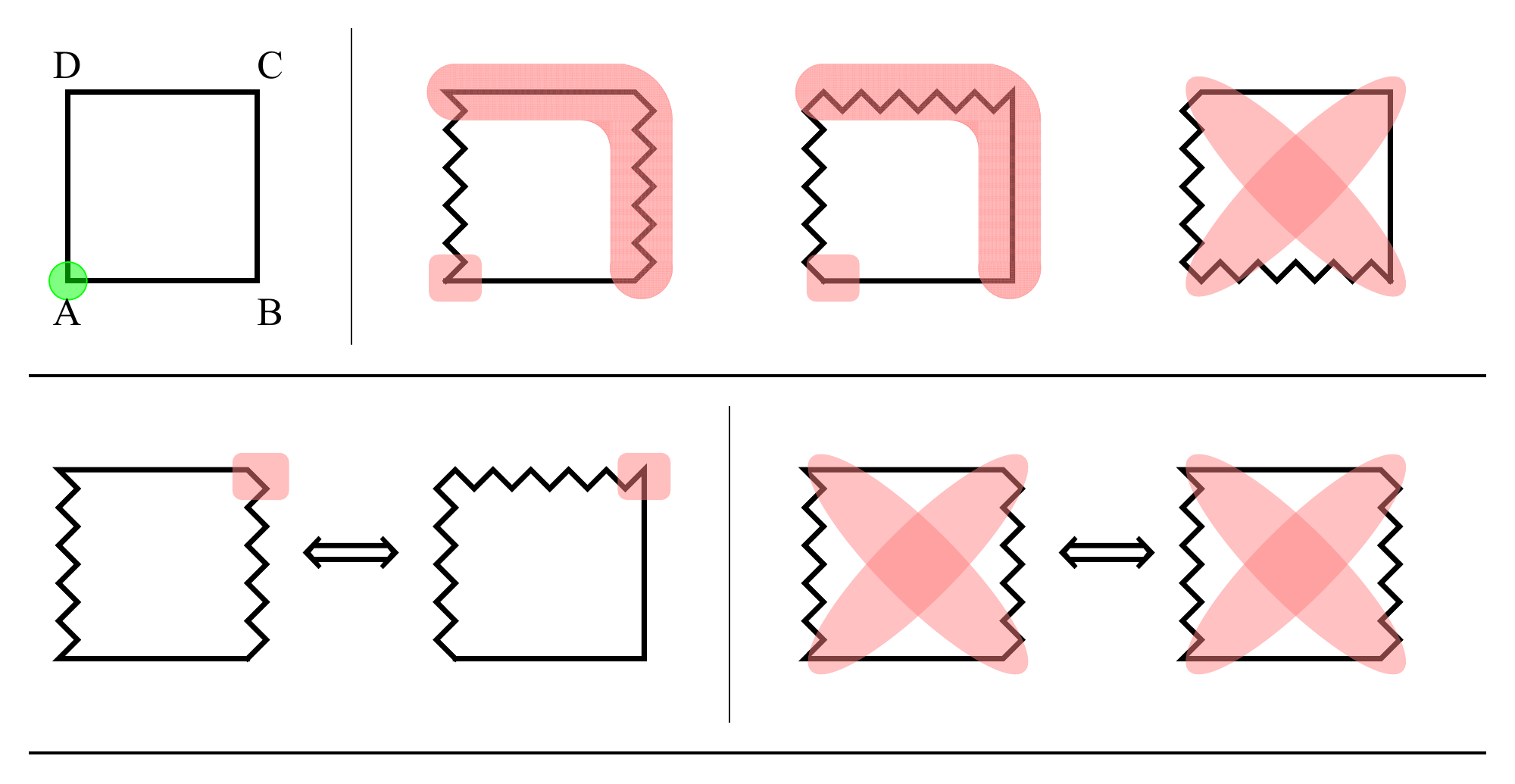}%
\caption{(Color online) Cluster decomposition rule for one of the motifs is shown in the figure. The premarked motifs for the gauge-term $J_0$ on the square lattice has four possible motifs corresponding to four sub-lattices of sites with single privileged site(pink rectangles). The zigzag lines indicate the bond consisting pair of anti-parallel spins, the frustrated bonds and unfrustrated bonds correspond to solid lines.}
\end{figure} 
where $A_{\square} = \sigma_{1}^{z}\sigma_{2}^{z}\sigma_{3}^{z}\sigma_{4}^{z}$, the indices $i$,$j$,$k$,$l$ represent the four spins on the edges of a plaquette, $B_{\square}=(\sigma_{1}^{z}\sigma_{3}^{z}$ + $\sigma_{2}^{z}\sigma_{4}^{z}$), the indices $i$,$j$ here represent spins opposite
to each other on a plaquette of the square lattice, and $P_{A,B}^x=\pm\frac{1}{4}(\sigma^z_{1}+\sigma^z_{2}-\sigma^z_{3}-\sigma^z_{4})$, $P_{A,B}^y=\pm\frac{1}{4}(\sigma^z_{1}+\sigma^z_{4}-\sigma^z_{2}-\sigma^z_{3})$ are the dipole-moment vectors for A,B sub-plaquettes as illustrated in Fig.(\ref{square})\\
\indent The parameter space of the Hamiltonian $H$ is three dimensional with $J_0$ being the largest followed by $K$, $J_1$, and $J_2$ in magnitude. $J_0$(>0) gives rise to $\mathcal{Z}_2$ gauge invariance and the intra-molecular coupling term $J_1$ gives rise to the restricted ice rules implying that only those states with ice rules having finite dipole moment are allowed. The presence of four-spin interactions signified by the coupling $J_0$ stems from the fact that for a real $\text{H}_2\text{SQ}$ molecule, the energy of a given molecule does depend on the relative position of the four hydrogen ions it shares with neighboring molecules~\cite{J.C.Slater}. This is particularly important because though at zeroth order the hydrogen ions are assumed to be tunneling in a quantum double well potential and thus an Ising-like two-spin interaction could also be a possible candidate for dynamics at zeroth order. However, as the height of the potential well is not significantly large, a more general four-spin interaction is more suitable~\cite{R.Blinc}. The second term of the Hamiltonian appears naturally from an approximated, possible two-spin interaction for a plaquette. Remember that for a plaquette there can be four nearest-neighbor Ising-like interactions and two Ising-like interactions across the diagonal links. The energy scale of these two types are different and the previous four types of Ising exchange interactions are an approximate representation of the general four-spin interaction represented by $J_0$. 
\indent Since there are ground states with each square plaquette having finite dipole moments, a more general Hamiltonian would also have these dipole-dipole interaction term $J_2$. The presence of dipole-dipole interactions is also inevitable from the fact that without their presence the observed ferroelectric transition would not be possible as argued in Ref. \citep{R.Blinc}. The physical origin of dipole-dipole interactions arises naturally when the local ground-state or zeroth-order electrostatic energy is disturbed by the vibrational modes of the "SQ" molecules thus giving rise to nearest-neighbor dipole-dipole interactions. 
\subsection{Stochastic series expansion quantum monte carlo method}
With the ever shifting focus on quantum monte carlo methods(QMC) for an unbiased calculations the hunt for an efficient cluster algorithms is on as far as stochastic series expansion(SSE) QMC method is concerned. The last few decades  has very limited success in developing efficient  algorithms at finite temperature especially for systems with frustration or for systems with macroscopic degeneracy. Earlier methods relied on constructing the clusters based on a "link-decomposition" of Hamiltonian found to be inefficient, despite being tried in various forms by constructing the clusters along $\tau$-dimension(imaginary-time). Nevertheless, here we use a variant of an recently developed microcanonical cluster algorithm(quantum cluster update) where it uses the plaquette decomposition of the Hamiltonian within the framework of Stochastic Series Expansion (SSE) pioneered by Sandvik\cite{Sand,Sandvik}.  Clusters are constructed  based on a plaquette percolation process with the notion of "premarked motifs"  which act as a flag in determining the way in which it connects the legs of all diagonal plaquette operators “living on”
various planes of imaginary-time direction. 
Once all clusters are constructed, each can be independently flipped with probability 1/2 within a  Swendsen-Wang  type implementation.   Here in the present study, we have employed and improvised an algorithm within the stochastic series expansion monte-carlo method\cite{Sounak}. Readers interested in the details regarding the efficiency and performance of the algorithm compared to percolation based algorithm can find here \cite{Sounak} for an "odd" Ising gauge Hamiltonian with anti-ferromagnetic Ising exchange term. 
The major difference in the design of the algorithm is that the choice of premarked motifs differ for each systems with complicated interactions or frustration, where an intuition or little prior knowledge of the equilibrium ensemble is a bonus for improving the algorithm further, this is also where we had implemented our idea to investigate the Hamiltonian we considered. \\
\indent  

Since there are three distinct regions corresponding to three different parameters $J_0, J_1, J_2$ with distinct equilibrium ensemble we thus identify different premarked motifs choice as shown in Fig.\eqref{cluster}. For $J_2=0$, we use the second choice of the premarked motifs(shown in third column of the Fig.\eqref{cluster}), as this choice naturally leads to the equilibrium ensemble of ground state manifold  where to minimize the Ising exchange interactions exactly two frustrated bonds(only those with one parallel to other) are required to be on each plaquette which corresponds to eigenvalue $4J_0+4J_1$, while configurations with four or zero frustrated bonds have eigen-value 0 and hence corresponding plaquette operators do not appear in the operator string. Our  choice  of  motif  in  the $J_2=0$  case  consists  of  two privileged diagonal sites  among  the  four  that  make  up  a spatial plaquette.  Thus each spatial plaquette has two distinct possible motifs. The  motif  on  a  given  spatial  plaquette  determines  the  cluster  decomposition  of  plaquette  operators at that location in the following way:  If only one frustrated bond touches the two privileged sites, the four legs corresponding to these two sites are assigned to a a priori different cluster, and the other four legs make up the other cluster.  If the privileged site is touched by two or zero frustrated bonds, then the four legs corresponding to the privileged site and its diagonally  opposite  site  are  assigned  to  one  cluster,  and the other four legs to an a priori different cluster (which could in principle merge with the other cluster at a future step in the cluster construction). \\
\indent We also study the parameter regime with $J_{1,2}=0$. Here the premarked motifs schemes pretty much remain same except that the number of choices to use motifs are six in the present case and hence can be invoked randomly on the fly. The premarked motif used in this case is shown in Fig.\eqref{cluster}. We see that on contrary to the case where $J_1 \neq 0$ there is no apparent spin-freezing in this case thus making the algorithm much more efficient in this regime. However, despite improvising the sophisticated algorithm to work in the case $J_1 \neq 0$ we see some apparent spin freezing at low temperatures and hence to avoid that we also invoke the replica exchange method with temperatures ranging from 0.05 to 1.0. We use system sizes varying from $L=24$ to $32$($N=2L^2$) with a standard $1\times10^7$ iterations for equilibrations and $1\times10^7$ for measurements. Results are divided into six bins to estimate statistical errors by the variance among the bins.

\begin{figure}[htp]
 \hspace{-3.2cm}
  \vspace{-1.5cm}
 \includegraphics[width=14.5cm,height=17.0cm,left]{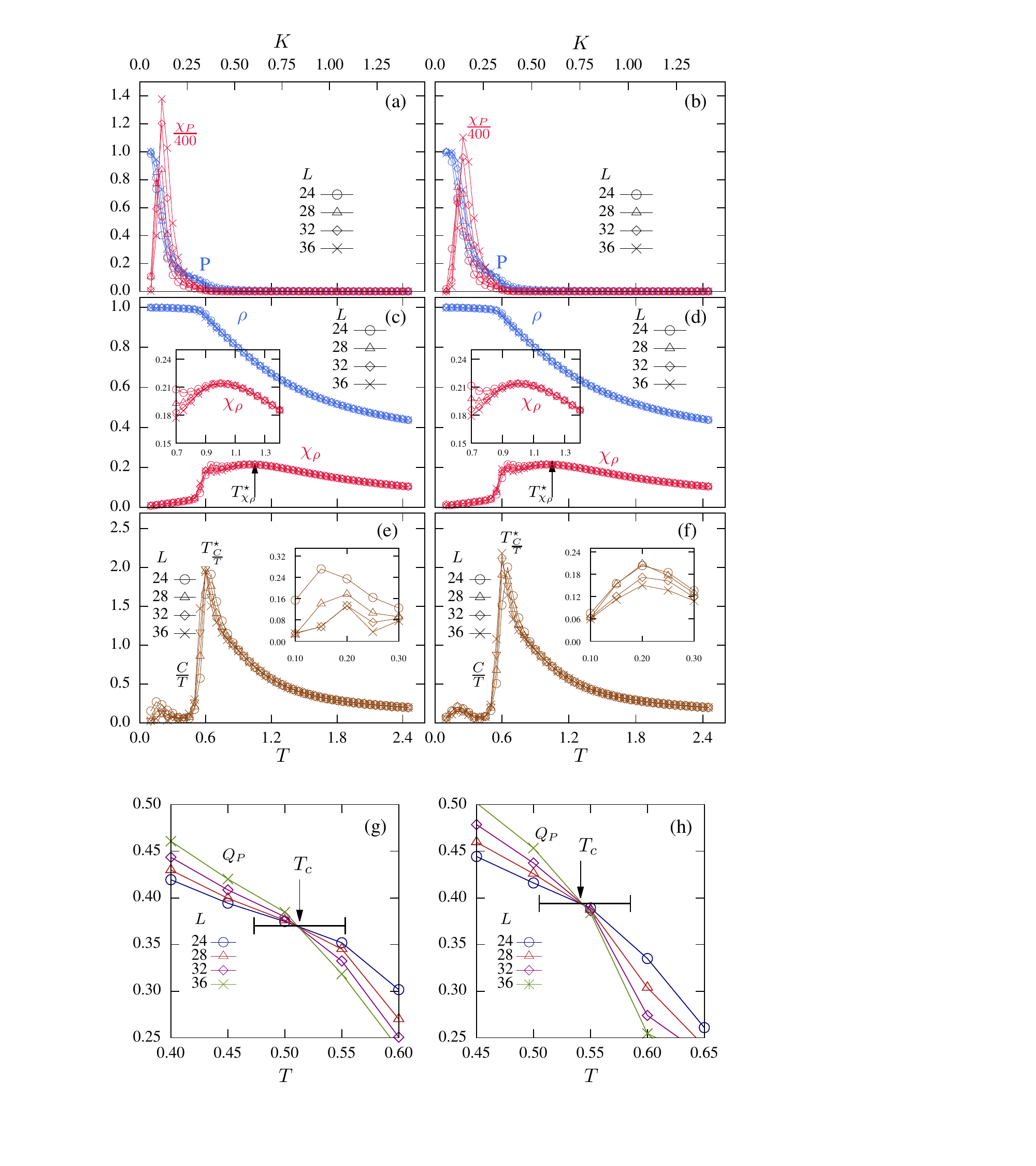}%
 \caption{(Color online)  In  panel (a) and (b) the red points and the blue points show the plot for $\chi_P$ and $P$. Similarly  in (c) and (d), we present $\rho$ and $\chi_{\rho}$ by 
red and blue points respectively. In (e),(f), specific heat divided by temperature($C/T$) has been shown.  Binder parameter has been shown in panel (g),(h). In all the plot, various 
different points denotes different system size as shown. All the results are calculated along the contour $K/T=\tan(\theta)$ where $\theta=\pi/6$. For detail description of the above,
we refer Sec. \ref{results-qmc}.  The parameter values used in the figures are $J_0=1.0$,$J_1=0.5$ with $J_2=0.020$ for left panel and $J_2=0.024$ for right panel.}
 \label{all_one}
 \end{figure}

\section{Physical quantities}
\label{phys-quant}
In view of the wide scope of our model which might host different phases dependent on the relative magnitude of the model parameter and temperature as well, we introduce the relevant parameters in detail.
This will help us  to identify each phases as well as to distinguish from each other without ambiguity.   At very high temperature $T >> J_i$, one generally expects a paraelectric
phases where the dipole-moments associated with each plaquettes are disordered. The paraelectric to ferroelectric phase transition (due to dipole-dipole interaction term $J_2$) as we lower the temperature is characterized by  the order parameter $P$ which is nothing but the electronic polarization and the associated susceptibility $\chi_P$. They are defined as below.
\begin{eqnarray}
P &=&\frac{1}{N}[\mid S(0,\pi)\mid^2 + \mid S(\pi,0)\mid^2]^{1/2} \\
\frac{\chi_P}{N}&=&\beta\bigg[\langle P^2\rangle - \langle P\rangle^2\bigg]
\end{eqnarray}
where S(\textbf{k}) is the static spin-structure factor given by
\begin{eqnarray}
S(\textbf{k})&=&\frac{1}{N}\sum\limits_{i,j}^{N} S_{i}^z S_{j}^z\exp{(-\textbf{k}\cdot\textbf{r}_{ij})} 
\end{eqnarray}

To locate the critical temperature associated with this paraelectric to ferroelectric phase transition, we use the Binder cumulant analysis, where 
the binder parameter $Q_P$ is defined as  

\begin{eqnarray}
\label{binder}
Q_P=\bigg(1-\frac{\langle P^4 \rangle}{8\langle P^2 \rangle^2}\bigg)
\end{eqnarray}

Apart from the presence of paraelectric and ferroelectric phases, at very low temperature and for small values of $J_2$, the strength of dipole-dipole interaction, the ground
state is dominated by states determined by $J_0$ and $J_1$. In this situation, the ground state manifold is dictated by the states which satisfy the "Ice rule". In true sense
this is a quantum liquid states. To distinguish this state from the usual paraelectric states, one need to define a order parameter which can successfully  establishes the
presence of this state and differentiate it from usual paraelectric phase and ferroelectric phase. To this end, following Ref.\cite{suzuki}, we define the parameter $\rho$ 
which detects the Ice-rule state, i.e, the partially disordered locally-correlated liquid-like paraelectric phase,
\begin{eqnarray}
\label{rho}
\rho=\frac{1}{N}\sum_{p}I(p)
\end{eqnarray}
where $I(p) \rightarrow 1$, if the given plaquette $p$ is in one of the local four-fold degenerate Ice-rule state and $I(p) \rightarrow -1/3$ otherwise. To further corroborate the results of phase 

transition and crossover we also calculate the susceptibility corresponding to  $\rho$ by calculating their fluctuations as given by,

\begin{eqnarray}
\frac{\chi_{\rho}}{N}&=&\beta\bigg[\langle \rho^2\rangle - \langle \rho\rangle^2\bigg] 
\end{eqnarray}


where the crossing over from the disordered dipoles into a paraelectric phase can as well be detected in the specific heat measurements as shown in Fig.\eqref{all_one}.

\begin{figure}[htp]
\hspace{-3.0cm}
\includegraphics[width=11.5cm,height=10.3cm,left]{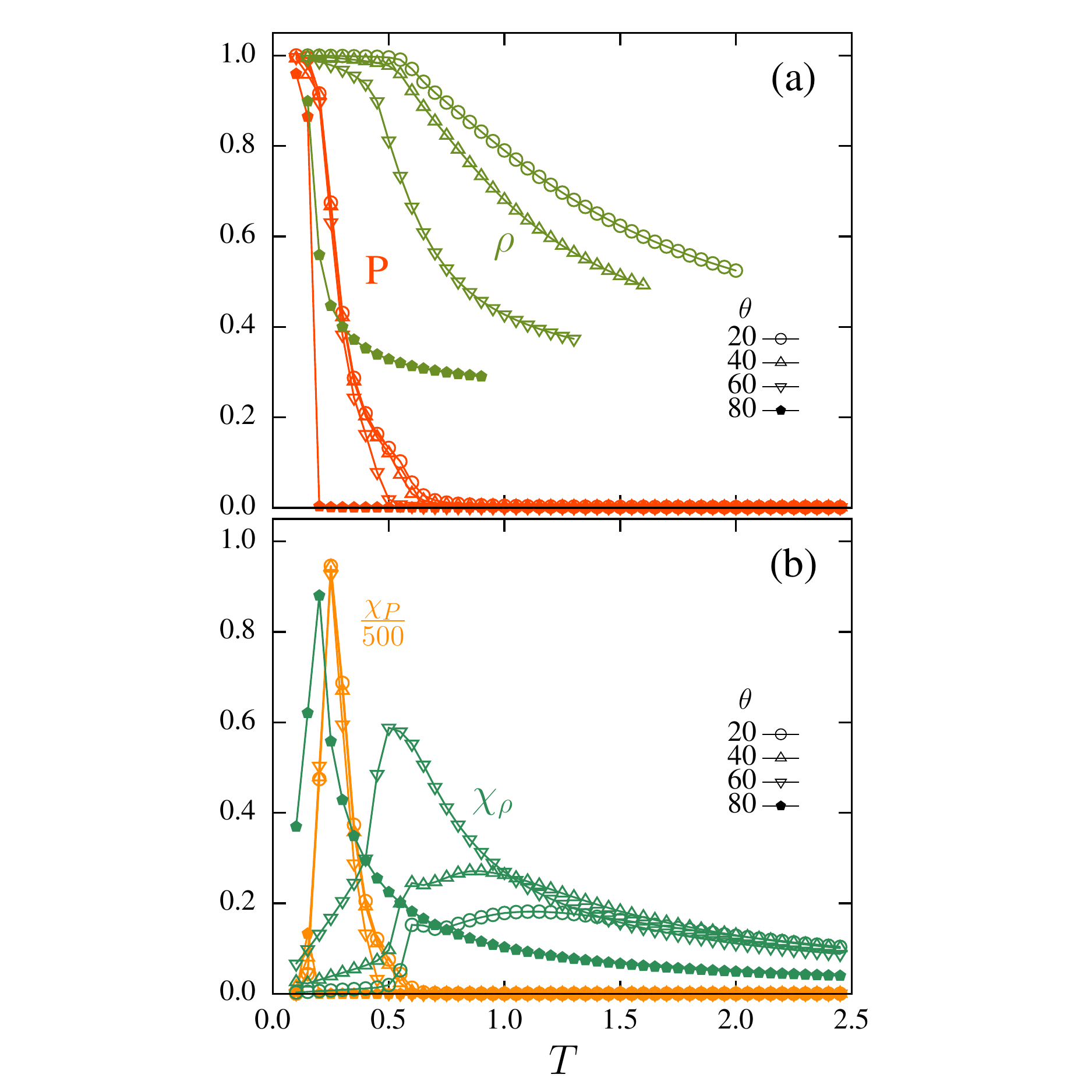}
\caption{(Color online) In the upper panel we show  temperature dependence of order parameter  and Ice-rule parameter $\rho$ by the orange and green points respectively. The corresponding susceptibility have been plotted in the lower panel with the same color convention as the upper plane. Note that in the above 2D plot each graph denotes a specific
line in the $T-K$ plane. Each line is represented by the corresponding slope of the line $\theta$. For the details of the plot kindly refer to the text in Sec. \ref{results-qmc}.}
\label{order-theta}
\end{figure}
 
\section{RESULTS}
\label{results-qmc}

We know from the earlier studies~\cite{Chern,Vikas} that presence of dipole-dipole interaction (the $J_2$ term in the Hamiltonian in Eq.\eqref{Hamiltonian} induces a ferroelectric order. In the absence of $J_2$, the ground state is highly degenerate. The degeneracy for $J_1=0$ is exponential though with $J_1$ it is proportional to the peripheral size of the system. On the
other hand the global ferroelectric order is four fold degenerate. Thus we see that the effect of temperature on the system might be very intricate due to the energy cost for low energy excitations due to competing interactions as well as due to degeneracy of the ground state manifold for each parameters. This actually necessitates to examine the order parameter $P$ and $\rho$ both. Interestingly
we find that they does not follow each other as we increase the temperature. Corresponding the susceptibility also shows  contrasting behaviour. We first discuss the temperature dependence of $P$ and
$\rho$ followed by the corresponding susceptibility.

\subsection{Temperature dependence of \texorpdfstring{$P \hspace{1mm}\text{and} \hspace{1mm} \rho$}{Lg}}

\indent The general behaviour as evident from the QMC simulation suggests that as we turn on the temperature the ferroelectric order parameter denoted by $P$ sharply decreases  for a very small value of temperature. However this decrease seems to be a  two steps process.  We call the first phase of decrease of $P$ as the quantum liquid states and the higher temperature counterpart which is the usual para electric phase.
The $P$ shows a shoulder like hump at the transition from quantum liquid like states to para electric states. In Fig. \eqref{all_one}(a-b), we have shown the variation of $P$ in blue points
for   $J_0=1.0,J_1=0.5,J_2=0.020,0.024$.  The presence of quantum liquid like states is apparent from the temperature variation of $\rho$ as presented in blue points in Fig. \eqref{all_one}(c-d).
We clearly observe that $\rho$ is almost constant throughout this quantum liquid like state and decreases monotonically when the system yields to paraelectric states. Thus the order parameter
$P$ and $\rho$ suggest that as we increase the temperature from  zero, the system starts from ferroelectric phase, moves to an intermediate  quantum-liquid like state and finally reaches to
a pare-electric phase. In the upper panel of Fig. \ref{order-theta} we have shown by orange and green points more plots for the behaviour of $P$ and  $\rho$ respectively for various values of $\theta$ in $T-K$plane.  It shows that for large values of $\theta$, $P$ and $\rho$ decreases more rapidly than the small values of $\theta$. This indicates that the pressure and temperature has opposite effect in the system.  The pressure denoted by $K$ tends to stabilize the ferromagnetic and the intermediate quantum liquid like states.

\subsection{Susceptibility \texorpdfstring{$\chi_P \hspace{1mm} \text{and} \chi_{\rho} \hspace{1mm} \text{and specific heat} \hspace{1mm} C/T$}{Lg}}

The susceptibility obtained due to $P$, is shown in Fig. \eqref{all_one}(a-b) by red points which shows a jump at the transition from ferroelectric phase to quantum liquid-like states. This suggests
that ferroelectricity is almost destroyed at this transition. However the susceptibility corresponding to $\rho$, i.e $\chi_{\rho}$ shows a very interesting feature. Initially it is almost zero, increases very slowly until the temperature reaches near to the transition from quantum liquid like states to the paraelectric states. At this transition the $\chi_{\rho}$ jumps at a higher value
and remain almost constant up to a certain temperature which we call $T_{\rho}$ and  after this, $\chi_{\rho}$ decreases monotonically. The specific heat at very large temperature shows monotonic decreasing  behaviour characteristic to usual para-electric phase but at low temperature it shows two peaks of different magnitude as denoted in Fig. \eqref{all_one}(e,f). The largest peak appears at the transition of quantum liquid like state and we denote this temperature by $T^*_{C/T}$. However, the sharp nature of the peak indicates a possible order-disorder phenomenon where the degeneracy seems to be uplifted to some extent. Below this temperature specific heat shows another small peak at  where the $P$ starts to decrease from initial constant value for small $T$.  The peak height of this smaller one tends to decrease with increase of system size as denoted in the inset of Fig. \eqref{all_one}. In the lower panel Fig. \ref{order-theta} we have shown by orange and green plots for the behaviour of $\chi_P$ and  $\chi_{\rho}$ respectively for various values of $\theta$ in $T-K$plane.  It shows that for large values of $\theta$, $P$ and $\rho$ has a more sharper peaks and also they decrease more rapidly compared to the small values of $\theta$. It suggests that the stability of the intermediate liquid like state is enhanced by increasing the $K/T$ ratio. This may be attributed to the fact that increase of $K$ results into enhancing the bandwidth of the system resulting into decreasing the thermal effect.

\subsection{Binder cumulant and critical temperature}

In the foregoing discussion we have already introduced the two critical temperature. The largest one is the $T^*_{\chi_{\rho}}$ which is signified by the step like jump from almost zero values  of $\chi_{\rho}$ to a higher value. Below this temperature we observed another critical temperature signified by the largest jump in the specific heat at the transition to paraelectric phase  for the quantum liquid-like state. This temperature is denoted by $T^*_{C/T}$. Our analysis for the Binder cumulant as shown in the Fig. \eqref{all_one} shows that the actual phase transition is very near to the $T^{*_{C/T}}$
and we  denote this temperature as $T_C$. As for as our  numerical results and analysis are concerned, $T_c=0.56(47)$ and $ 0.56(03)$ for $J_2=0.020$ and $0.030$ respectively. The corresponding $T^*_{C/T}$ is obtained as $0.060(03)$ and $0.61(40)$ respectively.  
\begin{figure}[htp]
\hspace{-1.0cm}
\vspace{-1.0cm}
\includegraphics[width=9.0cm,height=19cm,left]{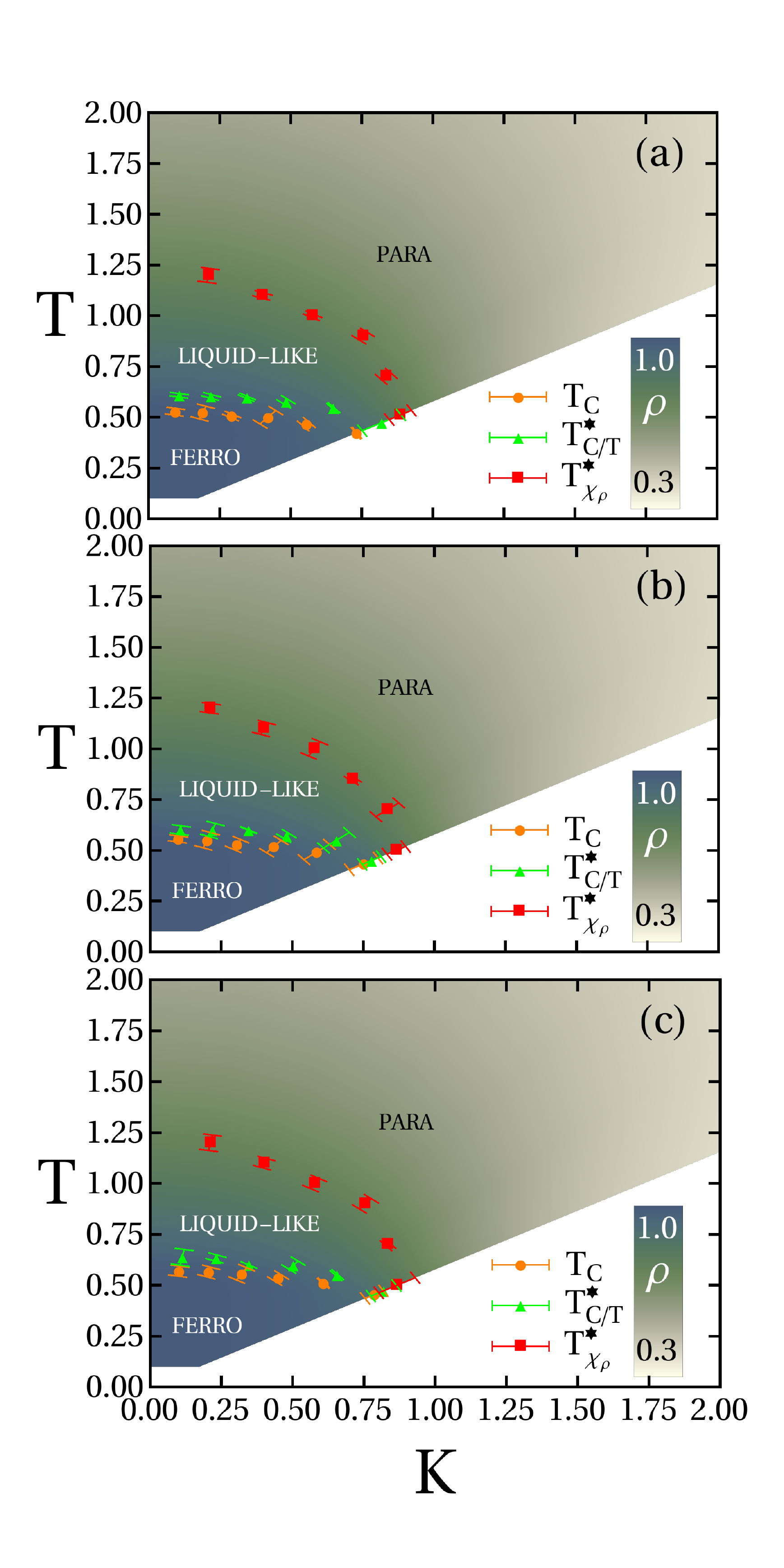}%
\caption{(Color online) Phase diagram showing the critical and crossover points, the figures (a)-(c) correspond to the values $J_2=0.020$,$J_2=0.024$,$J_2=0.028$ respectively. The orange dots indicate the critical points for ferro to liquid like transition obtained from the binder analysis, green triangles indicate the cross over points $T^{\star}_{C/T}$ obtained from the C/T curve, while the red squares indicate the $T^{\star}_{\chi_{\rho}}$ obtained from $\chi_{\rho}$.}
\label{phase}
\end{figure}

\subsection{Phase diagram}
The numerical results presented above suggests the presence of three phases as we increase the temperature. The first one is ferroelectric phase which survives for very small temperature and extends up to $T_c$ obtained from Binder cumulant. We call this phase as $\Pi_f$. The jump in the specific heat signified by $T_{C/T}$ is little higher than the $T_c$. 
Above the $T_c$ there is a presence of a complex quantum liquid like state which extends up to some critical temperature  $T^{\star}_{\chi_{\rho}}$. This phase is very intriguing
as long as the behaviour of order parameter is concerned and we call this phase as $\Pi_{ql}$. After $T^{\star}_{\chi_{\rho}}$, the normal para electric phase develops with no residual
quantum correlation between the dipole moments. The above three critical temperature defined above varies as the temperature and the $K$ is varied. It is instructive to present above three phases in the contour plot in $T-K$ plane and is presented in Fig.\eqref{phase}. The results are shown for different $J_2$ strengths in the $T-K$ plane. For the case where $J_2=0$ the system shows a crossover transition from quasi macroscopic degenerate liquid-like state to a completely disordered state, with the crossover temperatures shifting to higher temperatures with $J_1$ as expected. However even for $J_2$ very small, the system orders ferroelectrically(the region below $T_c$, orange circles), the region between the $T_c$ and $T^{\star}_{\chi_{\rho}}$ belongs to the liquid-like state. \\
\indent In contrast to the previous model, the scaling behaviour of the critical boundary shows interesting characteristics, the boundary doesn't extend into the liquid-like region rapidly but rather slowly indicating that the dependence of $T_c$ with $J_2$ is complex with lower dynamic exponent than the previous models\cite{suzuki}. Another aspect is that the ratio of thermal energy($T_c$) to that of $J_2$ is just 10 which is an order less compared to other model in hand\cite{suzuki}. We believe that the colossal enhancement of $T_c$ with $J_2$ is related to the lifting of quasi macroscopic degeneracy.  The above finding could be confirmed by experiment in conjunction with a first-principles study yielding an estimation of model Hamiltonian realistic parameter range. It is interesting to note that
all the phase transitions that we observe is not of first order but rather like second order specially the transition from $\Pi_{f}$ to $\Pi_{ql}$. On the other hand the transition from $ \Pi_{ql}$ to $\Pi_{p}$ is a smooth crossover. The phase $\Pi_{ql}$ and $\Pi_{f}$ suggests quite a interesting behaviour  of $\rho$ and $C/T$. Let us begin with $\Pi_{f}$ first. At $T=0$, the system
is in ferroelectric state with global ground state degeneracy of four. As we increase the temperature, the ferroelectric order parameter $P$ decreases rapidly but $\rho$ remains constant and in addition we observe small peak in $C/T$. This suggests that the global ferroelectric order reduces into domain of ferroelectric clusters and also includes the presence of other states not contributing
to $P$ but consistent with Ice rules. The possibility of presence of these additional states explains the small peaks in $C/T$ which also decreases as we increase the system size. As far as as the phase $\Pi_{ql}$ is concerned, we observe the steady decrease of the Ice rule order parameter $\rho$ which suggests that the excitations now includes the non-Ice rule states as  expected. However the quantum correlation is not lost completely until the critical temperature $T_{\chi_{\rho}}$ is reached.
\section{Discussion}
\label{disc}
We have considered a model where at zeroth order there is a four-spin Ising gauge like interaction of protons in the pseudo-spin formalism. Earlier results on the model showed that the system hosts a deconfined phase for $J_2=0$~\cite{Chern} at zero temperature. We use the model here to extend the phase diagram along the finite temperature axis. Our results show an Ice-rule dominated  strong proton-proton correlations to be the main physics of the system. Motivated by experiments and the previous studies, the  phase diagram in the $T-K$ plane suggests that the qualitative shape of the critical boundary $T_c$ is  more closer to  a linear behaviour for small field strength consistent with the experimental results\cite{E.Barth,W.Kuhn,M.Mehring}, though at large pressure,  $T_c$ and $T^{\star}$ does not behave linearly in contrast to experimental  results where one finds a complete linear behaviour for all ranges of applied pressure. Another aspect that we find in our study is the difference  between $T_c$ and $T^{\star}$ which remains constant through out all pressure ranges, which to our opinion a  remarkable success of our study.  However, our results indicate a possible second order phase transition in contrast to experiments.  This deviation  with experiments might be reconciled by considering the inter layer coupling or more complicated couplings to lattice distortions, which are neglected in our model. The temperature dependence of the interaction parameter can not be ruled out as well. The anomalous specific heat peak at low temperature in the ferroelectric phase is due to the formation of cluster of ferroelectric  domains. The height of the peak reduces with the increase of system size suggesting that  for small system, the domain size is comparable to the system itself.    \\ 
\indent   We note that a previous study~\cite{suzuki} has investigate the finite temperature phase diagram in a related model where  only two-spin interactions has been considered.  It is very important to compare the results and the conformity  and contrast between the two study is of quite academic interest.  We have found all the phases and the characteristic obtained in their study  for example, the phase transition from the ferroelectric phase to quantum liquid like states. However our results has improved many  aspects and seems more realistic as far as the experimental results are concerned.  For example, the nature of phase transition as denoted by the specific heat is more steeper. Similar characteristics is found in the behaviour of $\rho$ as well. This suggests that our results actually mimic the phase transition look like second order. 
Another important aspect is the nature of phase diagram in $T-K$ plane is linear for small and intermediate values of $K$ and for large values of $K$ it becomes elliptical
which was the case for all values of $K$ earlier\cite{suzuki}. We think the  distinct feature of our results shows that a four-spin interaction at the zeroth order is more realistic.
Also in the present model the dependence of degeneracy of ground state manifold is different from the two-spin Hamiltonian. For example,  at the zeroth order,  four spin interaction includes non-Ice rule states as well (all spin up and all spin down in a given plaquette) ice-rule states. The diagonal interaction reduces the  macroscopic ground state degeneracy to be proportional to linearly the system dimension. However for the two body case, the degeneracy is still macroscopic. We believe that this might be the reason that the present study is more closure to the experimental realization.  Another distinct feature is that, the ratio of thermal energy to the exchange energy of the dipole-dipole interaction $T_c/J_2$ is approximately 1/4 times the ratio obtained by earlier studies\cite{suzuki}, this peculiar aspect is ascribed to the fact that the ground state for $J_2=0$ is a strong deconfined phase robust to local perturbations induced in the form of dipole-dipole interaction and it may be the case where the underlying dynamics of the quasi-particle excitations in the deconfined phase could be independent of the dynamics of the excitations with that of the ferroelectric ground state.

\section{Summary}
\label{summ}

To summarize, we have studied the finite temperature phase diagram of proton dynamics of squaric acid system. The study offers a unique opportunity to examine
the competition between quantum fluctuation and thermal fluctuation and the effect of intricate ground state degeneracy of such system. The model Hamiltonian 
we considered involves a four-spin interaction which renders the present study an interesting one in view of the successfully solving the model
as well as to qualitatively reproduce the experimental findings.\\
\indent In our endeavour to study such a Hamiltonian ~\eqref{Hamiltonian} involving four-spin interaction, we used SSE quantum monte carlo involving an improvised version of loop update algorithm which efficiently overcome the problem of ergodic sampling in some parameter regimes. Our theoretical study successfully detected the quantum liquid-like intermediate state  before
the appearance of conventional paraelectric state as we increase the temperature. The ferroelectric state  which exist before the liquid like quantum liquid-like  state, is characterized by its  colossal dependence of $T_c$ with respect to $J_2$, the dipole-dipole interaction. In the intermediate liquid-like state, the local correlations governing the
ice-rule constraints are still valid to a large extent. While the experimental phase diagram in $T-K$ plane, the phase transition  shows a linear behaviour for all
 $ K $, we find the linear behaviour for small and intermediate values of $K$. Thus though our results fall short of full experimental confirmation, still it is
a quite an improvement in regard to the earlier theoretical study so far. Also the phase transition from the ferroelectric phase to quantum liquid like state is more
sharper which reasserts that the model which considered here is more practical as far as the squaric acid system is concerned. For further improvement of our results, a first-principle calculations would be helpful to establish the correct Hamiltonian as well as fixing the realistic parameters. In addition to this, the inter layer coupling should be considered
to fill the gap between the theoretical and experimental studies. This effort is beyond the scope of present study and kept as a future scope.

\section*{Acknowledgment}

The authors thank Kedar Damle and Sounak Biswas for fruitful discussions on SSE QMC. We also thank the support staff of High performance computing facility provided at Institute of Physics and funded by Homi Bhabha National Institute(HBNI).

\end{document}